\documentclass[12pt]{article}
\begin{document}
\pagestyle{empty}
\begin{flushright}
   hep-ph/0211016\\
   October 2002
\end{flushright}
\begin{center}
{\Large
{\bf Nambu-Goldstone Bosons in CP Violating theory\\
with Majorana Masses}}\\[1.cm]

{\large Darwin Chang$^{a,b}$, Wai-Yee Keung$^{b}$ and Chen-Pin
Yeh$^{a}$ }\\

{\em $^a$NCTS and Physics Department, National Tsinng Hua
University,\\ Hsinchu 30043, Taiwan, R.O.C.}\\

{\em $^b$Physics Department,University of Illinois at Chicago, IL
  60607-7059,  USA}\\
\end{center}

\bigskip\bigskip
\centerline{\small \it   ABSTRACT}
\vskip 1cm
We derive some properties of the Nambu-Goldstone boson coupling in
theories that have CP violation and Majorana masses. We show
explicitly that its diagonal coupling to a Majorana  fermion is
pseudoscalar not scalar. This clarifies some confusion in the
literature. Some potentially useful off-diagonal properties are
also derived. We also show, in the process, that the Goldstone
theorem often produces interesting and nontrivial identities in
matrix theory which may be hard to prove otherwise.
\\[0.3cm] PACS numbers: 14.80.Mz, 11.30.Er,  14.60.Pq

\bigskip\bigskip

\section*{Introduction}
Theories with Majorana masses and CP violation have been taken more
seriously recently because of the experimental evidence for
neutrino masses\cite{sno,sol,atm}. They often contain some
ingredients of spontaneous symmetry breaking of either gauge or
global symmetries, and  as a result, some accompanying unphysical
Higgs or Nambu-Goldstone (NG)  bosons.  For example, the theory with
spontaneously broken lepton number usually produces a NG
boson, usually called majoron. It was well known that the coupling
of the NG boson  to the fermion should be always
pseudoscalar in nature. This is because the degree of freedom of
the NG boson usually can be identified as the symmetry generator
which is spontaneously broken\cite{Abers:qs}. Under the symmetry
transformation, the associated NG boson is translated by a
constant.  For the theory to be invariant under this translation,
the NG boson has to couple only through its
derivative\cite{Chang:1988fz}.  For the couplings that are
diagonal in the fermion flavor, such derivative couplings
translate into pseudoscalar coupling using equation of motion.
However, there are some recent doubt on the applicability of the
above argument to the theory with Majorana masses and CP
violation\cite{He}.  In this paper we wish to address this issue
in detail and reaffirm that the above argument remains true.  The
conclusion is useful not only for the neutrino sector, but also
for the chargino and the neutralino sectors of the supersymmetric
theories\cite{mssm}.
For example, in Ref\cite{Chang:2002ex}, it was shown that the
diagonal couplings of the unphysical NG Higgs boson
to charginos are pseudoscalar even though the relevant
sector of the theory is CP violating  with Majorana gaugino
masses.

As a result of this analysis, we found that very interesting and
nontrivial theorems in matrix algebra can be derived as a direct
consequence of the Goldstone theorem.

In the next section we will use the simplest example in $SU(2)_L
\times U(1)$ context with one flavor of neutrino species
(including a right handed neutrino) to illustrate the pseudoscalar
nature of the NG boson coupling.  We will prove that the NG boson
coupling is pseudoscalar exactly.  Then we will generalize it to
$N$ flavors of neutrinos.  In both cases, the fact the diagonal
NG boson coupling has to be pseudoscalar gives rise to some
nontrivial matrix theorems.  To explore these matrix theorems
further, we consider a very general theory with many NG bosons and
then formulate a resulting matrix theorem which is simple but very
nontrivial.  In the appendix we include a general proof of the
pseudoscalar nature of the diagonal NG boson couplings for
completeness.

\section*{$2\times2$ matrices of
$\nu$ masses and couplings}
The model is a combination of the singlet majoron model\cite{sgltM}
and the triplet majoron model\cite{tpltM}.  Details of the model
can be found in Ref.\cite{He}. In addition to the known fermions
in the Standard Model (SM), there is an antineutrino (denoted as
$\nu^{c}$) which is a singlet under the gauge group with the
lepton number $L=-1$. The scalar bosons include one $SU(2)$
doublet $H$ (SM Higgs), one $SU(2)$ triplet $\chi$ and one gauge
singlet $S$.  $H$ has hypercharge $Y=-\frac12$, lepton number
$L=0$, and $\chi$ has $Y=1$, $L=-2$, $S$ has $Y=0$, $L=-2$.  Let
us start with one generation, the relevant quantum numbers are
summarized below 
$$ \begin{array} {|c|c|c|c|c|c|} \hline
    &   \nu      & \nu^c & H          & \chi & S  \\ \hline
Y   & -{1\over2} &  0    & -{1\over2} & 1    & 0  \\ \hline
L   & +1         & -1    & 0          & -2   & -2 \\ \hline
\end{array} \quad . $$
Redefining each scalar field as deviation from its
vacuum expectation value (vev), we write the Yukawa terms in the
Hamiltonian after symmetry breaking as
\begin{equation}
 \frac{2m_D}{v_H}\nu\nu^{c}(H^{0*}+v_H)
+\frac {M}{v_S}\nu^{c}\nu^{c}(S^*+v_S)
+\frac{m}{v_{\chi}}\nu\nu(\chi^0+v_{\chi})+\mbox{h.c.}
\end{equation}
Bilinear form of Weyl fermions are understood to be connected by
the Levi-Civita symbol. The neutrino mass terms are
\begin{equation}
\left(\!\!\begin{array}{cc} \nu & \nu^{c}
\end{array}\!\!\right) \left(\begin{array}{cc}
m&m_D\\
m_D&M             \end{array}\right)
\left(\!\!\begin{array}{c} \nu\\\nu^{c}
\end{array}\!\!\right)+\mbox{h.c.}
\end{equation}
Two of the three complex masses can be made real
by rotating the phases of the neutrino fields, leaving
{\it one} CP violating mass parameter.
The composition of the majoron can be derived from vacuum
expectation values as
\begin{equation}
G_M =\ (\sqrt{2}/N)
[ 2v_\chi^2v_H\hbox{Im}(H^0)
  +v_H^2v_\chi\hbox{Im}(\chi^0)
 +(v_H^2+4v_\chi^2)v_S\hbox{Im}(S)]   \ ,
\end{equation}
where
$N=\sqrt{4v_\chi^4v_H^2+v_H^4v_\chi^2+(v_H^2+4v_\chi^2)^2v_S^2}$.
The majoron-neutrinos coupling terms in the Hamiltonian are then
\begin{equation}
-\frac {i}{\sqrt{2} N}\left(\!\!\begin{array}{cc} \nu & \nu^c
\end{array}\!\!\right) \left(\begin{array}{cc}
-\frac m{v_{\chi}}v_{\chi}v_H^2&\frac{m_D}{v_H}2v_Hv_{\chi}^2\\
\frac{m_D}{v_H}2v_Hv_{\chi}^2&\frac M{v_S}v_S(v_H^2+4v_{\chi}^2)
\end{array}\right)
\left(\!\!\begin{array}{c} \nu\\\nu^c
\end{array}\!\!\right)G_M+\mbox{h.c.}
\end{equation}
The above coupling $2\times 2$ matrix has the simple form 
$ -i(\sqrt{2}N)^{-1} {\bf C}$, where ${\bf C}$ is
\begin{equation}
{\bf C}\equiv\left(\begin{array}{cc}
-mv_H^2&2m_Dv_{\chi}^2\\
2m_Dv_{\chi}^2&M(v_H^2+4v_{\chi}^2)
\end{array}\right)     \ .
\end{equation}
The symmetric neutrinos mass matrix $M_n$ can be diagonalized as
\begin{equation}
U^TM_nU=M_{\rm diag},~~~~M_n=\left(\begin{array}{cc}
m&m_D\\ m_D&M \end{array}\right) \ ,\label{eq:mn}
\end{equation}
where matrix $U$ is unitary and $M_{\rm diag}$ is a diagonal
matrix with real non-negative elements. In the mass eigenstate
basis the coupling matrix of the majoron to neutrinos becomes
$-i(\sqrt{2}N)^{-1} U^T{\bf C}U$.
The scalar couplings of the majoron are proportional to imaginary parts of
diagonal elements of $U^T{\bf C}U$.
One can show directly that this diagonal scalar coupling vanishes
by explicitly diagonalizing the mass matrix order by order in
$m/M$ (a procedure which was adopted in Ref\cite{He}). We will
give such an approximate calculation in the Appendix for
comparison with Ref\cite{He}.  Here we are going to  prove that the
diagonal scalar couplings of the majoron are zero using the matrix
property without assumption about the size of $m/M$.  Define
$r=2v_\chi^2/v_H^2$, we have
\begin{equation}
{\bf C}=v_H^2
\left(\begin{array}{cc}
-m&m_Dr\\ m_Dr&M(1+2r) \end{array}\right)
=v_H^2 (1+r)
\left(\begin{array}{cc} -m&0\\ 0&M\end{array}\right)
+v_H^2 r M_n
\ . \end{equation}
To prove that the diagonal elements of $U^T{\bf C}U$ are real,
we only need to show that
\begin{equation}
{\bf N}' \equiv U^T\left(\begin{array}{cc} -m&0\\ 0&M \end{array}\right)U
\end{equation}
has real diagonal elements.
\begin{eqnarray}
{\bf N'}_{11}&=&-mU_{11}^2+MU_{21}^2\\
{\bf N'}_{22}&=&-mU_{21}^2+MU_{22}^2    \ .
\end{eqnarray}
Denote $m_1, m_2$ as the neutrino mass eigenvalues. Then
\begin{equation}
\left(\begin{array}{cc}
m&m_D\\
m_D&M
\end{array}\right)=U^*\left(\begin{array}{cc}
m_1&0\\
0&m_2
\end{array}\right)U^{\dag}    \ .
\end{equation}
Express $m$ and $M$ as $m_1$ and $m_2$,
\begin{eqnarray}
m&=&(U_{11}^*)^2m_1+(U_{12}^*)^2m_2    \ ,\\
M&=&(U_{21}^*)^2m_1+(U_{22}^*)^2m_2    \ .
\end{eqnarray}
Unitarity of $U$ gives
\begin{eqnarray}
{\bf N'}_{11}&=&-((U_{11}^*)^2m_1+(U_{12}^*)^2m_2)U_{11}^2+((U_{21}^*)^2m_1+(U_{22}^*)^2m_2)U_{21}^2\nonumber\\
&=&((U_{21}U_{21}^*)^2-(U_{11}U_{11}^*)^2)m_1  \ ,\\
{\bf
N'}_{22}&=&-((U_{11}^*)^2m_1+(U_{12}^*)^2m_2)U_{12}^2+((U_{21}^*)^2m_1+(U_{22}^*)^2m_2)U_{22}^2\nonumber\\
&=&((U_{22}U_{22}^*)^2-(U_{12}U_{12}^*)^2)m_2   \ ,
\end{eqnarray}
which are real. So the diagonal scalar couplings of the majoron to
the neutrinos are zero even if the couplings are CP violating.

Next consider the off-diagonal terms of the majoron couplings.
Note that $U^T{\bf C}U$ and $v_H^2(1+r){\bf N}'$ have identical off-diagonal
components.  Using unitarity,
\begin{equation}
{\bf N'}_{12}={\bf N'}_{21}
= (m_1+m_2) \hbox{Re}(U_{21}^*U_{22})
+i(m_1-m_2) \hbox{Im}(U_{21}^*U_{22}) \ .
\end{equation}
It means, if the neutrinos are degenerate, the off-diagonal
couplings of the majoron are also pseudoscalar. If $(m_1+m_2)=0$
then the off diagonal couplings of the majoron are purely scalar.
This is not surprising because if $(m_1+m_2)=0$ then one can
redefine one of the neutrino by $i$, after that it is reduced to
the degenerate case and off-diagonal coupling picked up an $i$.
These results are consequences of the Goldstone Theorem.

If the global symmetry, say the lepton number in the current
example, is an accidental symmetry only valid for renormalizable
terms of fields in low energy physics, but broken by higher
dimensional operators induced from very short distance physics at
the scale $M_X$. In general, the NG boson picks up a tiny
mass\cite{tinymass} which is suppressed by the factor $1/M_X$.  CP
violation can probably give rise to a scalar diagonal coupling
which is also suppressed by $1/M_X$, not by the less reducing
factor $1/M$ from the neutrino sector. This mechanism\cite{fifthF}
provides a possible source of a feeble fifth force competing with
gravity.

\section*{$2N\times2N$ matrices of $N$ generations}
If there are $N$ generations of neutrinos. The above proof can
also be applied. We denote the $2N\times 2N$ symmetry mass matrix
as
\begin{equation}
M_n=\left(\begin{array}{cc}
\hat{m}&\hat{m_D}\\
\hat{m_D}^T&\hat{M}
\end{array}\right) \ ,
\end{equation}
where $\hat{m}$, $\hat{m_D}$ and $\hat{M}$ are $N\times N$ matrix.
We can also find a unitary matrix $U$ to diagonalize $M_n$ such
that
\begin{equation}
U^TM_nU=\left(\begin{array}{cc}
\hat{m}_1&0\\
0&\hat{m}_2
\end{array}\right)  \ ,\quad
U=\left(\begin{array}{cc}
\hat{U}_{11}&\hat{U}_{12}\\
\hat{U}_{21}&\hat{U}_{22}
\end{array}\right) \ ,
\end{equation}
where $\hat{m}_1$ and $\hat{m}_2$ are real $N\times N$ diagonal
matrices, and
$\hat{U}_{11},\hat{U}_{12},\hat{U}_{21},\hat{U}_{22}$
are four $N\times N$ matrix satisfying the
unitary conditions,
\begin{eqnarray}
\hat{U}^{\dag}_{11}\hat{U}_{11}+\hat{U}^{\dag}_{21}\hat{U}_{21}&=&{\bf1} \ ,
\label{eqn:unitary2}\\
\hat{U}^{\dag}_{12}\hat{U}_{12}+\hat{U}^{\dag}_{22}\hat{U}_{22}&=&{\bf1} \ ,\\
\hat{U}^{\dag}_{11}\hat{U}_{12}+\hat{U}^{\dag}_{21}\hat{U}_{22}&=&{\bf0} \ .
\label{eqn:unitary3}
\end{eqnarray}
The majoron coupling matrix in mass eigenstate basis
is given by
\begin{equation}
 -{i\over \sqrt{2}N} U^T\left(\begin{array}{cc}
-\hat{m}v_H^2&2\hat{m_D}v_{\chi^2}\\
2\hat{m_D}^Tv_{\chi}^2&\hat{M}(v_H^2+4v_{\chi}^2)
\end{array}\right)U   \ .
\end{equation}
Using the same trick as above,
we only need to show that
\begin{equation}
{\bf N'}=U^T\left(\begin{array}{cc}
-\hat{m}&0\\
0&\hat{M}
\end{array}\right)U
\end{equation}
has real diagonal elements.
First, we write down the 11 and 22 blocks,
\begin{eqnarray}
{\bf\hat{N}'}_{11}&=&
-\hat{U}_{11}^T\hat{m}\hat{U}_{11}+\hat{U}_{21}^T\hat{M}\hat{U}_{21} \ ,\\
{\bf\hat{N}'}_{22}&=&
-\hat{U}_{12}^T\hat{m}\hat{U}_{12}+\hat{U}_{22}^T\hat{M}\hat{U}_{22} \ .
\end{eqnarray}
We express $\hat{m}$ and $\hat{M}$ as
$\hat{m}_1$ and $\hat{m}_2$,
\begin{eqnarray}
\hat{m}&=&
 \hat{U}^*_{11}\hat{m}_1\hat{U}^{\dag}_{11}
+\hat{U}^*_{12}\hat{m}_2\hat{U}^{\dag}_{12}\ ,\\
 \hat{M}&=&
 \hat{U}^*_{21}\hat{m}_1\hat{U}^{\dag}_{21}
+\hat{U}^*_{22}\hat{m}_2\hat{U}^{\dag}_{22}\ .
\end{eqnarray}
Then, using the unitary conditions (\ref{eqn:unitary2}-\ref{eqn:unitary3}),
we obtain
\begin{eqnarray}
{\bf\hat{N}'}_{11}&=&
 -\hat{U}^T_{11}\hat{U}^*_{11} \hat{m}_1 \hat{U}^{\dag}_{11}\hat U_{11}
 +\hat{U}^T_{21}\hat{U}^*_{21} \hat{m}_1 \hat{U}^{\dag}_{21}\hat U_{21}
\nonumber\\
&=&\hat{m}_1- (\hat{m}_1 \hat{U}^{\dag}_{11}\hat{U}_{11}
                        +\hat{U}^T_{11}\hat{U}^*_{11}\hat{m}_1)
\ . \end{eqnarray}
Similarly,
\begin{equation}
{\bf\hat{N}'}_{22}=
 \hat{m}_2 \hat{U}^{\dag}_{22}\hat{U}_{22}
+          \hat{U}^{T}_{22}\hat{U}^*_{22}\hat{m}_2 -\hat{m}_2 \ .
\end{equation}
In terms of the mass eigenvalues $(\hat m_1)_i$ and $(\hat
m_2)_i$, we have
\begin{eqnarray}
({\bf\hat{N}'}_{11})_{ij}&=& +(\hat m_1)_i\delta_{ij}
- ((\hat m_1)_i +(\hat m_1)_j )\hbox{Re}(\hat U_{11}^\dagger \hat U_{11})_{ij}
-i((\hat m_1)_i -(\hat m_1)_j )\hbox{Im}(\hat U_{11}^\dagger \hat U_{11})_{ij}
 \ ,
\nonumber\\
({\bf\hat{N}'}_{22})_{ij}&=&  -(\hat m_2)_i \delta_{ij}
+  ((\hat m_2)_i +(\hat m_2)_j )\hbox{Re}(\hat U_{22}^\dagger \hat U_{22})_{ij}
+ i((\hat m_2)_i -(\hat m_2)_j )\hbox{Im}(\hat U_{22}^\dagger \hat U_{22})_{ij}
\ . \nonumber\\
\end{eqnarray}
No dummy index summation occurs in above expression.
It is obvious that the diagonal components of both
${\bf\hat{N}'}_{11}$ and ${\bf\hat{N}'}_{22}$ are real.
The off-diagonal block is
\begin{equation}
{\bf\hat{N}'}_{12}={\bf\hat{N}}^{'\ T}_{21}=
 \hat{m}_1 \hat{U}_{21}^{\dag}\hat{U}_{22}
+          \hat{U}_{21}^T\hat{U}_{22}^*\hat{m}_2
\end{equation}
\begin{equation}
({\bf\hat{N}'}_{12})_{ij} =
   ((\hat m_1)_i +(\hat m_2)_j )\hbox{Re}(U_{21}^\dagger U_{22})_{ij}
+ i((\hat m_1)_i -(\hat m_2)_j )\hbox{Im}(U_{21}^\dagger U_{22})_{ij} \ .
\end{equation}
These results all can be understood through arguments similar to
that in the last section.

\section*{General $N\times N$ mass matrix}
In this section, we demonstrate how the Goldstone theorem can be
used to prove a matrix theorem.\\
Theorem :
Given any  complex symmetric matrix of dimension $N \times N$,
\begin{equation}
M = \left(    \begin{array}{cccc}
m_{11}&m_{12}&\cdots&\\
m_{12}&m_{22}&&\\
\vdots&&\ddots&\\
&&&m_{NN}     \end{array}\right)\ ,
\end{equation}
we decompose $M$  as the sum of  ${1\over2}{\bf C}_i$,
\begin{equation}
{\bf C}_i=
\left(\begin{array}{ccccccc} \ddots&&&m_{1i}&&&\\
&0&&\vdots&&0&\\
&&\ddots&m_{i-1,i}&&&\\
&&&2m_{ii}&m_{i,i+1}&\cdots&m_{iN}\\
&&&&\ddots&&\\
&&&&&0&\\
&&&&&&\ddots
\end{array}\right)
\end{equation}
where ${\bf C}_i$ is nonzero only along the $i$'th row and $i$'th
column.
Let $U$ be the unitary matrix that diagonalizes the symmetric
complex matrix $M$, that is, $U^T M U=M_{D}$ where $M_{D}$ is a
real diagonal matrix. Then the diagonal elements of the matrix
$U^{T}{\bf C}U$ are also real, for  general real
linear combination of ${\bf C}_i$, that is,
\begin{equation}
{\bf C} =  \sum_{i=1}^N a_i {\bf C}_i =
\left(\begin{array}{cccccc}
2a_1m_{11} & (a_1+a_2) m_{12} &\cdots&(a_1+a_i)m_{1i}&\cdots&(a_1+a_N)m_{1N}\\
&2a_2m_{22}& \cdots & (a_2+a_i) m_{2i}&\cdots&  (a_2+a_N)m_{2N}\\
&&\ddots&\cdots&\cdots&\vdots\\
&&&2a_i m_{ii}&\cdots&(a_i+a_N)m_{iN}\\
&&&&\ddots&\vdots\\
&&&&&a_N m_{NN}
\end{array}\right) \ ,
\end{equation}
with arbitrary real coefficients $a_1,\cdots, a_{N}$.

We also know that if two of eigenvalues of $M_{D}$ are equal, say
$(M_{D})_{ii}=(M_{D})_{jj}$, then the correspondent term $(U^T{\bf
C}U)_{ij}$ is real. If $(M_{D})_{ii}=-(M_{D})_{jj}$, then the
correspondent term $(U^T{\bf C}U)_{ij}$ is pure imaginary.

The proof of this seemingly very nontrivial theorem follows
immediately from the Goldstone theorem.  Consider a Lagrangian
with $U_1(1)\times U_2(1)\cdots\times U_N(1)$ symmetry. For each
$U_i(1)$, there are one Weyl fermion $\psi_i$ which carries a unit
of its quantum number and one scalar $H_i$ with $-2$ units of its
quantum number.  In addition,  for each pair of $(U_i(1),U_j(1))$ there is
one scalar, $h_{ij}$ carrying the quantum number $(-1,-1)$.
There are ${1\over2}N(N-1)$ independent $h_{ij}$
as $h_{ij} \equiv h_{ji}$ and $i\neq j$.
The Lagrangian of Yukawa couplings can be written as
\begin{equation}
{\cal L}=\sum^N_{i=1}g_i\psi_i\psi_iH_i
+\sum^N_{\stackrel{i,j=1}{i\neq j}}f_{ij}\psi_i\psi_jh_{ij}~~+~~\mbox{h.c.}
\end{equation}
where $f_{ij}=f_{ji}$.  The couplings can be complex.  After
symmetry breaking,  scalars develop   vev's, which are real
when phases are absorbed into couplings with redefinition of fields
($\langle H_i   \rangle =v_i,
  \langle h_{ij}\rangle =v_{ij}$).
We define $m_{ii}=g_iv_i$,
$m_{ij}=f_{ij}v_{ij}$ (no summation). The resulting complex
symmetric Majorana mass matrix is exactly in the form of $M$
above.
The spectrum has $N$ NG bosons
\begin{equation}
G_i=-\frac{\sqrt{2}}{N_i}(2v_i\hbox{Im}(H_i)+\sum^N_{\stackrel{j=1}{j\neq
i}}v_{ij}\hbox{Im}(h_{ij}))
\end{equation}
where
$N_i=\sqrt{4v_i^2+\sum^N_{\stackrel{j=1}{j\neq i}}v^2_{ij}}$.
The Yukawa coupling of $G_i$ is
$-i(\sqrt{2}N)^{-1} {\bf C}_i$.
The conclusion follows immediately from the Goldstone theorem.
The explicit proof is very similar to those in the last two sections.
Basically, we need to show ${\bf C}_i-M$ after mass diagonalization
has  real diagonal entries.
Note that the simple example of the one generation model described by
Eq.~(1) corresponds to the present case of $N=2$ with $U_1(1)$
generated by the quantum number $Y$, and $U_2(1)$ generated by $Y+L/2$.

\section*{Conclusion}

In this paper, we basically reaffirm that the diagonal Goldstone boson
coupling should be pseudoscalar even when there are CP violation and
Majorana masses in the system.  Such a property is probably known,
however, it is still quite complicated to work it out
explicitly considering that there exists confusion in the
literature\cite{He}.  The pseudoscalar character of the unphysical
Higgs couplings to chargino and neutralinos were not worked out
explicitly until recently\cite{Chang:2002ex}.  This nontrivial character
implies that the Goldstone theorem often results in matrix theorems
which are very nontrivial unless one uses the Goldstone theorem to
motivate it.

\section*{Acknowledgment}

WYK is partially supported by a grant from U.S. Department of
Energy (Grant No. DE-FG02-84ER40173). DC is supported by a grant
from National Science Council(NSC) of Republic of China (Taiwan).
We wish to thank X.G. He and H. Haber for discussions.
DC wish to thank theory groups at SLAC and LBL for hospitality
during his visit.
WYK wish to thank NCTS of NSC for support during his sabbatical leave.
\vskip .5cm
\hrule
\vskip .5cm
\centerline{\huge  Appendix}
\vskip .5cm

\section*{Approximation for $2\times 2$ Majoron Model}

Below we work out the detailed calculation of the case studied in
Ref.\cite{He} for the neutrino mass matrix of size  $2\times2$ of
one generation. Physical CP phase $\phi$ appears in $M=|M|e^{i\phi}$
of Eq.~(\ref{eq:mn}) in the basis that $m$ and $m_D$ are real.
We assume $|M|\gg m_D\gg m > 0 $ and  ${m_D^2}/|M|$  the same
order as $m$.

First, we rotate phases of the two diagonal elements of $M_n$ in
opposite directions so that the resulting imaginary parts are
equal. Then, leaving temporarily the unit matrix with imaginary coefficient,
we easily diagonalize the remaining real symmetric matrix.
Putting back the imaginary part, we
make the final phase rotations to obtain the physical real diagonal
mass matrix. These steps are summarized as
$$ U\simeq
\left(\begin{array}{cc} e^{i{\phi\over2}}& 0 \\
                        0                & e^{-i{\phi\over2}}\end{array}\right)
\left(\begin{array}{cc} 1-im\sin\phi/|M  & 0 \\
                        0                & 1+im\sin\phi/|M|          \end{array}\right)
$$
$$ \qquad\qquad\qquad \times
\left(\begin{array}{cc} 1       & m_D/|M| \\
                       -m_D/|M| & 1       \end{array}\right)
\left(\begin{array}{cc} e^{-{i\over2}\theta} & 0 \\
                        0                & 1-im\sin\phi/|M| \end{array}\right)
\ , $$
with $ e^{i\theta} = m\cos\phi-m_D^2/|M| + i m\sin\phi $.
We find that
\begin{equation}
U \simeq \left(\begin{array}{cc}
  (1-im\sin\phi/|M|)e^{-{i\over2}(\theta-\phi)} & m_D/|M| e^{i\frac{\phi}{2}}\\
 - m_D/|M|  e^{-{i\over2}(\theta+\phi)}   & e^{-i\frac{\phi}{2}}
\end{array}\right) \ ,
\end{equation}
\begin{equation}M_{\rm diag}\simeq\left(\begin{array}{cc}
m_1  & 0\\ 0 & |M| \end{array}\right) \ ,\quad\hbox{ with }
m_1=\sqrt{m^2\sin^2\phi +(m\cos\phi-m_D^2/|M|)^2} \ .
\end{equation}
Denote the two neutrino mass eigenstates as $\psi_l$ and $\psi_h$.
The majoron-neutrinos coupling terms become
\begin{equation}
-\frac {i}{\sqrt{2}N}
\left(\!\!\begin{array}{cc}
\psi_l&\psi_h\end{array}\!\!\right)U^T
{\bf C}
U \left(\!\!\begin{array}{c} \psi_l\\\psi_h
\end{array}\!\!\right)G_M+\mbox{h.c.}
\end{equation}
\begin{equation}
\hbox{with }\qquad
{\bf C}=  \left(\begin{array}{cc}
               -mv_H^2    &           2m_Dv_{\chi}^2\\
           2m_Dv_{\chi}^2 & |M|e^{i\phi}(v_H^2+4v_{\chi}^2)
  \end{array}\right) \  .
\end{equation}
Following the usual procedure,
we pair up each left-handed Weyl field $\psi$ with its complex
conjugated field $\epsilon \psi^\dagger$  to form a four-component
Dirac field $\Psi$.
In term of the heavy and light Dirac fields $\Psi_l,\Psi_h$, the
interaction is described by
\begin{equation}
\frac{1}{\sqrt{2}N}
(\bar\Psi_l \quad \bar\Psi_h)[ \hbox{Im}(U^T{\bf C}U)
                          +\hbox{Re}(U^T{\bf C}U)i\gamma_5]
\left(\begin{array}{c}\Psi_l \\ \Psi_h \end{array}\right) G_M   \ .
\end{equation}
In the leading order, we obtain
\begin{equation}
(U^T{\bf C}U)_{11}=-v_H^2 m_1
\ ,\quad
(U^T{\bf C}U)_{22}=(v_H^2+4v_{\chi}^2)|M| \ ,
\end{equation}
with no imaginary parts to the accuracy of our expansion. This
confirms that the diagonal scalar couplings of the majoron are
zero, contrary to the conclusion in Eq.(16) of Ref\cite{He}. 

\section*{General Proof of Pseudoscalar Nature}
The pseudoscalar nature of the Goldstone boson when coupled to
the fermion is very generic, not limited to the neutrino models
discussed above. It applies to the Goldstone boson coupling to
chargino and neutralino as well. Below, we work out the detail in the
basis of $\psi_R$ and $\psi_L$.

The fermion mass matrix $M$ defined in the weak
basis is diagonalized by bi-unitary transformation,
\begin{equation} U' M U^\dagger = M_D  \ ,\qquad
-{\cal L}_M \supset \bar\psi_R M \psi_L
                 = \bar\Psi_R M_D \Psi_L \ .\end{equation}
Different choices of vev's correspond to various rotations
$e^{iT\phi}$ generated by the generator $T$ of the spontaneously
broken symmetry.  With the overall physics unchanged, the physical
diagonal mass matrix $M_D$ maintains basis independent, if we impose
the following substitutions,
\begin{equation} \begin{array}{rcl}
   U^\dagger &\to&   e^{ i\phi T_L}U^\dagger        \ ,      \\
   U'        &\to& U'e^{-i\phi T_R}  \ .\end{array} \end{equation}
The invariance $\delta\bigl(U' M U^\dagger\bigr)=0$ gives
\begin{equation}U'(\delta M) U^\dagger
=-[ (\delta U') MU^\dagger + U'M(\delta U^\dagger) ]
=-[ (\delta U') {U'}^\dagger M_D
              + M_D(U\delta U^\dagger) ]   \ .\end{equation}
For an infinitesimal $\phi$, we have
\begin{equation}  \delta M= i\phi (Tv)_i \hbox{$d\over dv_i$} M  \ ,\quad
  \delta U'       =-i\phi U'T_R  \ ,\quad
  \delta U^\dagger= i\phi T_LU^\dagger  \ . \end{equation}
Therefore,
\begin{equation}   Y \equiv [ U' T_R {U'}^\dagger M_D - M_D(UT_L U^\dagger) ]   \ ,\quad
     U'(Tv_i)(\hbox{$d\over dv_i$}M) U^\dagger = -Y\ .\end{equation}
As $M_D$ is diagonal and $U' T_R {U'}^\dagger$, $UT_L U^\dagger$ are
Hermitean, the diagonal entries of above expression $Y$ is purely
real.

The NG boson corresponding to the generator $T$ is
\begin{equation} G=(Tv)_i \hbox{Im} \sqrt{2}\Phi_i/N \ , \hbox{ or, }
\sqrt{2}\hbox{Im}\Phi_i =(Tv)_i G/N +\cdots \ , \end{equation}
with the normalization $N=|T v|$.
Here we adopt the convention  that $v_i$ are all real.
This can always be achieved by field
redefintion, sometimes inducing  complex couplings.
Its Yukawa terms in the Lagrangian become
\begin{equation} {\cal L}\supset
 - \bar\Psi_R U' (i\hbox{Im}\Phi_i \hbox{$d\over dv_i$}M)U^\dagger \Psi_L
=-i (\sqrt{2}N)^{-1} G\bar\Psi_R U' (Tv)_i (\hbox{$d\over dv_i$}M)
          U^\dagger \Psi_L \ , \end{equation}
\begin{equation} {\cal L}\supset
(\sqrt{2}N)^{-1} G[ i \bar\Psi_R Y        \Psi_L
                   -i \bar\Psi_L Y^\dagger\Psi_R]
= -(\sqrt{2}N)^{-1}G  \bar\Psi_i Y_{ii} i\gamma_5  \Psi_i +\cdots \ . \end{equation}
This proves the diagonal coupling of NG boson is pseudoscalar without
assuming CP conservation.

In the following, we provide another way to understand this result by
the current conservation.
The divergence of the $T$ current is zero because of symmetry.
\begin{equation} J^{T,\mu} = \bar\psi_L T_L \gamma^\mu \psi_L
              +\bar\psi_R T_R \gamma^\mu \psi_R
+ \Phi^\dagger i\stackrel{\leftrightarrow}{\partial^\mu}T \Phi  \ ,\end{equation}
\begin{equation} J^{T,\mu} = \bar\Psi_L \gamma^\mu U  T_L  U^\dagger \Psi_L
              +\bar\Psi_R \gamma^\mu U' T_R {U'}^\dagger\Psi_R
          - 2   (N/\sqrt{2}) \partial^\mu G +\cdots \ .
\ . \end{equation}
In the low energy limit,
\begin{equation}  \partial_\mu[\bar\Psi_L \gamma^\mu  U  T_LU^\dagger \Psi_L
 +\bar\Psi_R \gamma^\mu  U' T_R {U'}^\dagger\Psi_R]=
 (\sqrt{2}N) \partial_\mu\partial^\mu G
\ . \end{equation}
Making use of the equation of motion,
$i\not\!\partial \Psi_{L,R}=m\Psi_{R,L}$, we obtain
\begin{equation}(\sqrt{2}N) \partial_\mu\partial^\mu G = -i
      \bar\Psi_R [U' T_R {U'}^\dagger M_D
                 -M_D(UT_L U^\dagger) ]\Psi_L+\hbox{H.c.}  \ .\end{equation}
\begin{equation} (\sqrt{2}N)\partial_\mu\partial^\mu G =  -
    [ i \bar\Psi_R Y        \Psi_L
    -i \bar\Psi_L Y^\dagger\Psi_R]   \end{equation}
Thus, this equation can be effectively generated by
\begin{equation} {\cal L}_{\rm eff.} =
\hbox{$1\over2$}
(\partial G)^2 + \bar\Psi i\not\!\partial \Psi
- [\bar\Psi_R M_D \Psi_L +\hbox{H.c.}]
- (\sqrt{2}N)^{-1}G[ i \bar\Psi_R Y \Psi_L -i \bar\Psi_L Y^\dagger\Psi_R]   \ .\end{equation}
So we have consistent results from two different approaches.
\end{document}